\documentclass[12pt]{article}
  \usepackage{amsfonts}
  \usepackage{amsmath}
\usepackage{amssymb}
\usepackage{amscd}
\usepackage{graphicx}
\begin{document}

~~
\bigskip
\bigskip
\begin{center}
{\Large {\bf{{{Black-body radiation for twist-deformed space-time}}}}}
\end{center}
\bigskip
\bigskip
\bigskip
\begin{center}
{{\large ${\rm {Marcin\;Daszkiewicz}}$}}
\end{center}
\bigskip
\begin{center}
\bigskip

{ ${\rm{Institute\; of\; Theoretical\; Physics}}$}

{ ${\rm{ University\; of\; Wroclaw\; pl.\; Maxa\; Borna\; 9,\;
50-206\; Wroclaw,\; Poland}}$}

{ ${\rm{ e-mail:\; marcin@ift.uni.wroc.pl}}$}

\end{center}
\bigskip
\bigskip
\bigskip
\bigskip
\bigskip
\bigskip
\bigskip
\bigskip
\bigskip
\begin{abstract}
In this article we formally investigate the impact of twisted space-time on black-body radiation phenomena, i.e. we derive the $\theta$-deformed Planck distribution function as well as we
perform its numerical integration to the $\theta$-deformed total radiation energy. In such a way we indicate that the space-time noncommutativity very strongly damps the black-body radiation process. Besides we provide for small parameter $\theta$ the twisted counterparts of
 Rayleigh-Jeans and Wien distributions respectively.
\end{abstract}
\bigskip
\bigskip
\bigskip
\bigskip
\eject

\section{{{Introduction}}}

The suggestion to use noncommutative coordinates goes back to
Heisenberg and was firstly  formalized by Snyder in \cite{snyder}.
Recently, there were also found formal  arguments based mainly  on
Quantum Gravity \cite{2}, \cite{2a} and String Theory models
\cite{recent}, \cite{string1}, indicating that space-time at Planck
scale  should be noncommutative, i.e., it should  have a quantum
nature. Consequently, there are a number of papers dealing with
noncommutative classical and quantum  mechanics (see e.g.
\cite{mech}-\cite{qmgnat}) as well as with field theoretical models
(see e.g. \cite{prefield}-\cite{fiorewess}) in which  the quantum
space-time is employed.

It is well-known that  a proper modification of the Poincare and
Galilei Hopf algebras can be realized in the framework of Quantum
Groups \cite{qg1}, \cite{qg3}. Hence, in accordance with the
Hopf-algebraic classification  of all deformations of relativistic
and nonrelativistic symmetries (see \cite{class1}, \cite{class2}),
one can distinguish three
types of quantum spaces \cite{class1}, \cite{class2} (for details see also \cite{nnh}):\\
\\
{ \bf 1)} Canonical ($\theta^{\mu\nu}$-deformed) type of quantum space \cite{oeckl}-\cite{dasz1}
\begin{equation}
[\;{ x}_{\mu},{ x}_{\nu}\;] = i\theta_{\mu\nu}\;, \label{noncomm}
\end{equation}
\\
{ \bf 2)} Lie-algebraic modification of classical space-time \cite{dasz1}-\cite{lie1}
\begin{equation}
[\;{ x}_{\mu},{ x}_{\nu}\;] = i\theta_{\mu\nu}^{\rho}{ x}_{\rho}\;,
\label{noncomm1}
\end{equation}
and\\
\\
{ \bf 3)} Quadratic deformation of Minkowski and Galilei  spaces \cite{dasz1}, \cite{lie1}-\cite{paolo}
\begin{equation}
[\;{ x}_{\mu},{ x}_{\nu}\;] = i\theta_{\mu\nu}^{\rho\tau}{
x}_{\rho}{ x}_{\tau}\;, \label{noncomm2}
\end{equation}
with coefficients $\theta_{\mu\nu}$, $\theta_{\mu\nu}^{\rho}$ and  $\theta_{\mu\nu}^{\rho\tau}$ being constants.\\
\\
Moreover, it has been demonstrated in \cite{nnh}, that in the case of the
so-called N-enlarged Newton-Hooke Hopf algebras
$\,{\mathcal U}^{(N)}_0({ NH}_{\pm})$ the twist deformation
provides the new  space-time noncommutativity of the
form\footnote{$x_0 = ct$.},\footnote{ The discussed space-times have been  defined as the quantum
representation spaces, so-called Hopf modules (see e.g. \cite{oeckl}, \cite{chi}), for the quantum N-enlarged
Newton-Hooke Hopf algebras.}
\begin{equation}
{ \bf 4)}\;\;\;\;\;\;\;\;\;[\;t,{ x}_{i}\;] = 0\;\;\;,\;\;\; [\;{ x}_{i},{ x}_{j}\;] = 
if_{\pm}\left(\frac{t}{\tau}\right)\theta_{ij}(x)
\;, \label{nhspace}
\end{equation}
with time-dependent  functions
$$f_+\left(\frac{t}{\tau}\right) =
f\left(\sinh\left(\frac{t}{\tau}\right),\cosh\left(\frac{t}{\tau}\right)\right)\;\;\;,\;\;\;
f_-\left(\frac{t}{\tau}\right) =
f\left(\sin\left(\frac{t}{\tau}\right),\cos\left(\frac{t}{\tau}\right)\right)\;,$$
$\theta_{ij}(x) \sim \theta_{ij} = {\rm const}$ or
$\theta_{ij}(x) \sim \theta_{ij}^{k}x_k$ and  $\tau$ denoting the time scale parameter
 -  the cosmological constant. Besides, it should be  noted that the mentioned above quantum spaces {\bf 1)}, { \bf 2)} and { \bf 3)}
 can be obtained  by the proper contraction limit  of the commutation relations { \bf 4)}\footnote{Such a result indicates that the twisted N-enlarged Newton-Hooke Hopf algebra plays a role of the most general type of quantum group deformation at nonrelativistic level.}.

Recently, there has been discussed the impact of different  kinds of
quantum spaces on the dynamical structure of physical systems (see e.g. \cite{mech}-\cite{field} and \cite{romero}-\cite{giri}).
Particulary, it has been demonstrated, that
in the case of a classical oscillator model \cite{kijanka} as well as in the case of a nonrelativistic particle moving in constant
external field force $\vec{F}$ \cite{daszwal}, there are generated by space-time noncommutativity additional force terms. Such a type
of investigation has been  performed for quantum oscillator model as well \cite{kijanka}, i.e., it was demonstrated that the  quantum space in nontrivial way affects  the spectrum of the energy operator. Besides, in the paper \cite{toporzelek} there has been considered a model of a particle moving on the $\kappa$-Galilei space-time in the presence of gravitational field force. It has been demonstrated, that in such a case
there is produced a force term, which can be identified with the so-called Pioneer anomaly \cite{pioneer}, and the value of the deformation parameter $\kappa$ can be fixed by a comparison of obtained result with observational data.
Moreover,  the quite interesting results have been obtained in the series of papers
\cite{hallcan}-\cite{hallcanf} concerning the Hall effect for canonically deformed space-time (\ref{noncomm}). Particularly, there has been found the $\theta$-dependent (Landau)
energy spectrum of an electron moving in uniform magnetic as well as in uniform electric field. Such results have been generalized to the case
of the twisted N-enlarged Newton-Hooke Hopf algebra in paper \cite{daszlandau} and \cite{daszlandau1}. Finally, it should be mentioned that similar  investigation has been performed in the context of black-body radiation process as well. For example, it has been demonstrated with use of  noncommutative electromagnetic field theory that black-body radiation for quantum space becomes anisotropic. A direct implication of such a result on Cosmic Microwave Background map has been argued in papers \cite{bb1} and \cite{bb2}.

In this article we also investigate the impact of twisted space-time on black-body radiation phenomena. However we assume
that a single mode of photon field oscillates with frequency predicted by new, first-quantized and canonically noncommutative oscillator model.
Precisely we derive the $\theta$-deformed Planck distribution function as well as we
perform its numerical integration to the $\theta$-deformed total radiation energy. In such a way we indicate that the space-time noncommutativity  very strongly damps the black-body radiation process. Besides we provide for small parameter $\theta$ the twisted counterparts of
 Rayleigh-Jeans and Wien distributions respectively.

The paper is organized as follows. In Sect. 2 we recall basic facts
concerning the most wide class of twisted (N-enlarged Newton-Hooke)   space-times
\cite{nnh} which includes the canonically deformed one as well. The third section is devoted to the calculation of isotropic energy spectrum  for the oscillator model
defined on such quantum spaces. In section four the Planck distribution function is provided and its numerical integration to the $\theta$-deformed total radiation energy is performed. The
final remarks are presented in the last section.

\section{Twisted N-enlarged Newton-Hooke space-times}

In this section we recall the basic facts associated with the twisted N-enlarged Newton-Hooke Hopf algebra $\;{\cal U}^{(N)}_{\alpha}(NH_{\pm})$ and with the
corresponding quantum space-times \cite{nnh}.  Firstly, it should be noted that in accordance with Drinfeld  twist procedure, the algebraic sector of twisted
Hopf structure $\;{\cal U}^{(N)}_{\alpha}(NH_{\pm})$ remains
undeformed, i.e., it takes the form
 \begin{eqnarray}
&&\left[\, M_{ij},M_{kl}\,\right] =i\left( \delta
_{il}\,M_{jk}-\delta _{jl}\,M_{ik}+\delta _{jk}M_{il}-\delta
_{ik}M_{jl}\right)\;\; \;, \;\;\; \left[\, H,M_{ij}\,\right] =0
 \;,  \label{q1} \\
&~~&  \cr &&\left[\, M_{ij},G_k^{(n)}\,\right] =i\left( \delta
_{jk}\,G_i^{(n)}-\delta _{ik}\,G_j^{(n)}\right)\;\; \;, \;\;\;\left[
\,G_i^{(n)},G_j^{(m)}\,\right] =0 \;,\label{q2}
\\
&~~&  \cr &&\left[ \,G_i^{(k)},H\,\right] =-ikG_i^{(k-1)}\;\; \;, \;\;\; \left[\, H,G_i^{(0)}\,\right] =\pm \frac{i}{\tau}G_i^{(1)}\;\;\;;\;\;\;k>1
\;,\label{q3}
\end{eqnarray}
where $\tau$, $M_{ij}$, $H$, $G_i^{(0)} (=P_i)$, $G_i^{(1)} (=K_i)$ and $G_i^{(n)} (n>1)$ can be identified with cosmological time parameter, rotation, time translation, momentum, boost and accelerations  operators respectively. Besides, the coproducts and antipodes of considered algebra are given by\footnote{$\Delta_0(a) = a\otimes 1 +1\otimes a$, $S_{0}(a) =-a$.}
\begin{equation}
 \Delta _{\alpha }(a) = \mathcal{F}_{\alpha }\circ
\,\Delta _{0}(a)\,\circ \mathcal{F}_{\alpha }^{-1}\;\;\;,\;\;\;
S_{\alpha}(a) =u_{\alpha }\,S_{0}(a)\,u^{-1}_{\alpha }\;,\label{fs}
\end{equation}
with $u_{\alpha }=\sum f_{(1)}S_0(f_{(2)})$ (we use the Sweedler's notation
$\mathcal{F}_{\alpha }=\sum f_{(1)}\otimes f_{(2)}$) and with the twist factor
$\mathcal{F}_{\alpha } \in {\cal U}^{(N)}_{\alpha}(NH_{\pm}) \otimes
{\cal U}^{(N)}_{\alpha}(NH_{\pm})$
satisfying  the classical cocycle condition
\begin{equation}
{\mathcal F}_{{\alpha }12} \cdot(\Delta_{0} \otimes 1) ~{\cal
F}_{\alpha } = {\mathcal F}_{{\alpha }23} \cdot(1\otimes \Delta_{0})
~{\mathcal F}_{{\alpha }}\;, \label{cocyclef}
\end{equation}
and the normalization condition
\begin{equation}
(\epsilon \otimes 1)~{\cal F}_{{\alpha }} = (1 \otimes
\epsilon)~{\cal F}_{{\alpha }} = 1\;, \label{normalizationhh}
\end{equation}
such that ${\cal F}_{{\alpha }12} = {\cal F}_{{\alpha }}\otimes 1$ and
${\cal F}_{{\alpha }23} = 1 \otimes {\cal F}_{{\alpha }}$.

The corresponding quantum space-times are defined as the representation spaces (Hopf modules) for the N-enlarged Newton-Hooke Hopf algebra
\;${\cal U}_{\alpha}^{(N)}(NH_{\pm})$. Generally, they are equipped with two the spatial directions
commuting to classical time, i.e. they take  the form
\begin{equation}
[\;t,\hat{x}_{i}\;] =[\;\hat{x}_{1},\hat{x}_{3}\;] = [\;\hat{x}_{2},\hat{x}_{3}\;] =
0\;\;\;,\;\;\; [\;\hat{x}_{1},\hat{x}_{2}\;] =
if({t})\;\;;\;\;i=1,2,3
\;. \label{spaces}
\end{equation}
 However, it should be noted
that this type of noncommutativity  has  been  constructed explicitly  only in the case of the 6-enlarged Newton-Hooke Hopf algebra, with
\cite{nnh}\footnote{$\kappa_a = \alpha$ $(a=1,...,36)$ denote the deformation parameters.}
\begin{eqnarray}
f({t})&=&f_{\kappa_1}({t}) =
f_{\pm,\kappa_1}\left(\frac{t}{\tau}\right) = \kappa_1\,C_{\pm}^2
\left(\frac{t}{\tau}\right)\;, \nonumber\\
f({t})&=&f_{\kappa_2}({t}) =
f_{\pm,\kappa_2}\left(\frac{t}{\tau}\right) =\kappa_2\tau\, C_{\pm}
\left(\frac{t}{\tau}\right)S_{\pm} \left(\frac{t}{\tau}\right) \;,
\nonumber\\
&~~&~~~~~~~~~~~~~~~~~~~~~~~~~~~~~~~~~ \nonumber\\
&~~&~~~~~~~~~~~~~~~~~~~~~~~~~~~~~~~~~\cdot \nonumber\\
&~~&~~~~~~~~~~~~~~~~~~~~~~~~~~~~~~~~~\cdot \label{w2}\\
&~~&~~~~~~~~~~~~~~~~~~~~~~~~~~~~~~~~~\cdot \nonumber\\
&~~&~~~~~~~~~~~~~~~~~~~~~~~~~~~~~~~~~ \nonumber\\
f({t})&=&
f_{\kappa_{35}}\left(\frac{t}{\tau}\right) = 86400\kappa_{35}\,\tau^{11}
\left(\pm C_{\pm} \left(\frac{t}{\tau}\right)  \mp \frac{1}{24}\left(\frac{t}{\tau}\right)^4 - \frac{1}{2}
\left(\frac{t}{\tau}\right)^2 \mp 1\right) \,\times \nonumber\\
&~~&~~~~~~~~~~~~~~~~\times~\;\left(S_{\pm} \left(\frac{t}{\tau}\right)  \mp \frac{1}{6}\left(\frac{t}{\tau}\right)^3 - \frac{t}{\tau}\right)\;,
\nonumber\\
f({t})&=&
f_{\kappa_{36}}\left(\frac{t}{\tau}\right) =
518400\kappa_{36}\,\tau^{12}\left(\pm C_{\pm} \left(\frac{t}{\tau}\right)  \mp \frac{1}{24}\left(\frac{t}{\tau}\right)^4 - \frac{1}{2}
\left(\frac{t}{\tau}\right)^2 \mp 1\right)^2\;, \nonumber
\end{eqnarray}
and
$$C_{+/-} \left(\frac{t}{\tau}\right) = \cosh/\cos \left(\frac{t}{\tau}\right)\;\;\;{\rm and}\;\;\;
S_{+/-} \left(\frac{t}{\tau}\right) = \sinh/\sin
\left(\frac{t}{\tau}\right) \;.$$
Moreover, one can easily check that when $\tau$ is approaching the infinity limit the above quantum spaces reproduce the canonical (\ref{noncomm}),
Lie-algebraic (\ref{noncomm1}) and quadratic (\ref{noncomm2})  type of
space-time noncommutativity, i.e., for $\tau \to \infty$ we get
\begin{eqnarray}
f_{\kappa_1}({t}) &=& \kappa_1\;,\nonumber\\
f_{\kappa_2}({t}) &=& \kappa_2\,t\;,\nonumber\\
&\cdot& \nonumber\\
&\cdot& \label{qqw2}\\
&\cdot& \nonumber\\
f_{\kappa_{35}}({t}) &=& \kappa_{35}\,t^{11}\;, \nonumber\\
f_{\kappa_{36}}({t}) &=& \kappa_{36}\,t^{12}\;. \nonumber
\end{eqnarray}
Of course, for all deformation parameters $\kappa_a$  going to zero the above deformations disappear.

\section{Quantum oscillator model for twisted N-enlarged Newton-Hooke space-time}

Let us now turn to the oscillator model defined on quantum space-times (\ref{spaces})-(\ref{qqw2}).
In first step of our construction, we extend the described in pervious section spaces to the whole algebra of momentum and position operators as follows
\begin{eqnarray}
&&[\;\hat{ x}_{1},\hat{ x}_{2}\;] = 2if_{\kappa_a}({t})\;\;\;,\;\;\;
[\;\hat{ p}_{1},\hat{ p}_{2}\;] = 2ig_{\kappa_a}({t})\;,\label{rel1}\\
&&[\;\hat{ x}_{i},\hat{ p}_{j}\;] =
i\hbar \delta_{ij}\left[1+ f_{\kappa_a}({t})g_{\kappa_a}({t})/\hbar^2\right]\;, \label{rel2}
\end{eqnarray}
with the arbitrary function $g_{\kappa_a}({t})$. One can check that relations
(\ref{rel1}), (\ref{rel2}) satisfy the Jacobi identity and for deformation parameters
$\kappa_a$ approaching zero become classical. \\
Next, by analogy to the commutative case, we define the Hamiltonian operator
\begin{eqnarray}
\hat{H} = \frac{1}{2m}\left({\hat{{p}}_1^2}+{\hat{{p}}_2^2} \right) +
\frac{1}{2}m\omega^2 \left({\hat{{x}}_1^2}+{\hat{{x}}_2^2} \right) \;. \label{2dhn}
\end{eqnarray}
with $m$ and $\omega$ denoting the mass and frequency of a particle, respectively. \\
In order to analyze the above system, we represent the
noncommutative operators $({\hat x}_i, {\hat p}_i)$ by the classical
ones $({ x}_i, { p}_i)$ as  (see e.g.
\cite{romero1}-\cite{kijanka})
\begin{eqnarray}
{\hat x}_{1} &=& { x}_{1} - {f_{\kappa_a}(t)}p_2/\hbar\;,\label{rep1}\\
{\hat x}_{2} &=& { x}_{2} +{f_{\kappa_a}(t)}p_1/\hbar
\;,\label{rep2}\\
{\hat p}_{1} &=& { p}_{1} + {g_{\kappa_a}(t)}x_2/\hbar\;,\label{rep2a}\\
{\hat p}_{2} &=& { p}_{2} -{g_{\kappa_a}(t)}x_1/\hbar
\;,\label{rep3}
\end{eqnarray}
where
\begin{equation}
[\;x_i,x_j\;] = 0 =[\;p_i,p_j\;]\;\;\;,\;\;\; [\;x_i,p_j\;]
={i\hbar}\delta_{ij}\;. \label{classpoisson}
\end{equation}
Then, the  Hamiltonian (\ref{2dhn}) takes the form
\begin{eqnarray}
\hat{H} = \hat{H}(t) =
\frac{1}{2M(t)}\left({{{p}}_1^2}+{{{p}}_2^2} \right)  +
\frac{1}{2}M(t)\Omega^2(t)\left({{{x}}_1^2}+{{{x}}_2^2} \right)
- S(t)L\;, \label{2dh1}
\end{eqnarray}
with
\begin{eqnarray}
&&L = x_1p_2 - x_2p_1\;, \\
&&1/M(t) = 1/m +m\omega^2 f_{\kappa_a}^2(t)/\hbar^2 \;,\\
&&\Omega(t) = \sqrt{\left(1/m
+m\omega^2 f_{\kappa_a}^2(t)/\hbar^2 \right)\left(m\omega^2
+ g_{\kappa_a}^2(t)/(\hbar^2m)\right)}\;,
\end{eqnarray}
and
\begin{equation}
S(t)=m\omega^2f_{\kappa_a}(t)/\hbar +g_{\kappa_a}(t)/(\hbar m)\;.
\end{equation}

 In accordance with the scheme proposed in  \cite{kijanka},  we introduce a set of time-dependent creation
$(a^{\dag}_{A}(t))$ and annihilation
$(a_{A}(t))$ operators 
\begin{eqnarray}
\hat{a}_{\pm}(t) &=& \frac{1}{2}\left[\frac{({p}_2 \pm
i{p}_1)}{\sqrt{M(t)\Omega (t)\hbar}} -i\sqrt{M(t)\Omega
(t)/\hbar}({x}_2 \pm
i{x}_1)\right]\;,\label{oscy1}
\end{eqnarray}
satisfying the standard commutation relations
\begin{eqnarray}
[\;\hat{a}_{A},\hat{a}_{B}\;] =
0\;\;,\;\;[\;\hat{a}^{\dag}_{A},\hat{a}^{\dag}_{B}\;]
=0\;\;,\;\;[\;\hat{a}_{A},\hat{a}^{\dag}_{B}\;] =
\delta_{AB}\;\;\;;\;\;A,B = \pm\;.\label{ccr1}
\end{eqnarray}
Then, it is easy to see  that in  terms of the  objects (\ref{oscy1}) the Hamiltonian function (\ref{2dh1}) can be written as
follows
\begin{equation}
{\hat{H}}(t)=\hbar\Omega_{+}(t) \left[{\hat N}_+(t) + \frac{1}{2}\right]
+ \hbar\Omega_{-}(t) \left[{\hat N}_-(t) + \frac{1}{2}\right]  \;, \label{hamquantosc}
\end{equation}
with the  coefficient  $\Omega_{\pm}(t)$ and the particle number operators $\hat{N}_{\pm}(t)$
given by
\begin{eqnarray}
\Omega_{\pm}(t)&=&\Omega(t)\mp S(t)\;,\label{ompm}\\
{\hat N}_{\pm}(t)&=&{\hat a}^{\dag}_{\pm}(t){\hat
a}_{\pm}(t)\;.\label{nn}
\end{eqnarray}
Besides, one can observe that
the eigenvectors of Hamiltonian (\ref{hamquantosc}) can be written as
\begin{eqnarray}
|n_+,n_-,t> =
\frac{1}{\sqrt{n_+!}}\frac{1}{\sqrt{n_-!}}\left({\hat
a}^{\dag}_{+}(t)\right)^{n_+} \left({\hat
a}^{\dag}_{-}(t)\right)^{n_-}|0>\;,\label{state}
\end{eqnarray}
while the corresponding eigenvalues take the form
\begin{equation}
E_{n_+,n_-}(t) = \hbar\Omega_{+}(t) \left[n_+ + \frac{1}{2}\right] +
\hbar\Omega_{-}(t) \left[n_- + \frac{1}{2}\right]\;. \label{eigenvalues}
\end{equation}

Let us now consider an interesting situation such that
\begin{eqnarray}
S(t) = 0\;.\label{equality}
\end{eqnarray}
One can check that it appears  when functions $f_{\kappa_a}(t)$ and $g_{\kappa_a}(t)$ satisfy the following condition
\begin{eqnarray}
f_{\kappa_a}(t) = -{g_{\kappa_a}(t)}/({\omega^2m^2}) \;.\label{condition}
\end{eqnarray}
Then, we have
\begin{eqnarray}
\Omega_{-}(t)=\Omega_{+}(t)=\Omega(t) =  m\omega\left({1}/{m}
+{m\omega^2} f_{\kappa_a}^2(t)/{\hbar^2}\right)
\;,\label{mamy}
\end{eqnarray}
and, consequently, the spectrum (\ref{eigenvalues}) provides the energy levels for two-dimensional isotropic oscillator model with the time-dependent frequency
$\Omega(t)$
\begin{equation}
E_{n_+,n_-}(t) = \hbar\Omega(t) \left[n_+ + \frac{1}{2}\right] +
\hbar\Omega(t) \left[n_- + \frac{1}{2}\right] \;. \label{landau}
\end{equation}
Particularly,
for canonical deformation $f_{\kappa_1}(t) = \kappa_1 = \theta$ we get
\begin{equation}
E_{n_+,n_-,\theta} = \hbar\Omega_{\theta} \left[n_+ + \frac{1}{2}\right] +
\hbar\Omega_{\theta} \left[n_- + \frac{1}{2}\right] \;, \label{landauek}
\end{equation}
with constant frequency
\begin{equation}
\Omega_{\theta} = \Omega_{\theta}(\omega) = m\omega\left({1}/{m}
+{m\omega^2}\theta^2/{\hbar^2}\right)
\;, \label{stalacze}
\end{equation}
such that $\lim_{\theta \to 0}\Omega_{\theta} = \omega$.

\section{Black-body radiation for twisted space-time}

In this section we derive the black-body radiation for $\theta$-deformed nonrelativistic space, i.e. due to the results of pervious section we
take under consideration the following energies of a single mode of photon field\footnote{Due to the isotropy of spectrum (\ref{landauek}) we consider excitations only in one direction. Besides as a single (emitted)
quanta we take $E=E_{n+1}-E_n = \hbar \Omega_{\theta}(\omega)$.}
\begin{equation}
E_{n,\theta}(\omega) = \hbar \Omega_{\theta}(\omega) n \;\;\;;\;\;\;n = 1,2,3,\; \ldots
\;, \label{modaen}
\end{equation}
where factor $\Omega_{\theta}(\omega)$ is given just by (\ref{stalacze}). Consequently, in accordance with quantum theory \cite{planck}
its average energy takes the form
\begin{equation}
\overline{E}_{\theta}(\omega) = \frac{\hbar \Omega_{\theta}(\omega)}{\exp\left(\frac{\hbar\Omega_{\theta}(\omega)}{kT}\right)-1}
\;, \label{sredniaen}
\end{equation}
with Boltzman constant $k$ and temperature $T$.  Besides, due to the fact that the number of state with frequency from $\omega$ to $\omega + d\omega$ per volume
unit is \cite{planck}, \cite{huang}
\begin{equation}
\rho(\omega)d\omega = \frac{8\pi\omega^2}{c^3}d\omega
\;, \label{perunit}
\end{equation}
we get the following $\theta$-deformed Planck distribution function\footnote{The distribution function $f_{\theta}(\omega)$ has been plotted
for different values of parameters $\theta$ and $T$ on Figure 1 and 2 respectively.}
\begin{equation}
f_{\theta}(\omega) = \frac{8\pi\hbar\omega^2}{c^3}\cdot\frac{\Omega_{\theta}(\omega)}{\exp\left(\frac{\hbar\Omega_{\theta}(\omega)}{kT}\right)-1}
\;. \label{planckfunction}
\end{equation}
Of course, in  $\theta$ approaching zero limit we reproduce from (\ref{planckfunction}) well-known Planck
formula
\begin{equation}
f(\omega) = \frac{8\pi\hbar}{c^3}\cdot\frac{\omega^3}{\exp\left(\frac{\hbar\omega}{kT}\right)-1}
\;, \label{nonplanckfunction}
\end{equation}
while for small values of deformation  parameter, we have
\begin{eqnarray}
f_{\theta}(\omega) &=& \frac{8\pi\hbar}{c^3}\cdot\frac{\omega^3}{\exp\left(\frac{\hbar\omega}{kT}\right)-1} \;+\nonumber\\
&-&\frac{8\pi\hbar}{c^3}\cdot\frac{m^2 \omega^5 \left(\hbar\omega\exp\left(\frac{\hbar\omega}{kT}\right)
- k T \left(\exp \left(\frac{\hbar\omega}{kT}\right)-1\right)\right) \theta^2}{\left(\exp\left(\frac{\hbar\omega}{kT}\right)-1\right)^2 \hbar^2 k
T}  \;+\label{smallplanckfunction} \\
&+& {\cal O}(\theta^3)\;. \nonumber
\end{eqnarray}
Besides in accordance with (\ref{smallplanckfunction}) in high-temperature $(kT>>\hbar\omega)$ as well as in high-frequency $(\hbar\omega >> kT)$ limit, we obtain the
$\theta$-deformed counterparts of Rayleigh-Jeans
\begin{eqnarray}
f_{\theta}(\omega) &=& \frac{8\pi\omega^2\hbar^2kT}{c^3} - \frac{8\pi m^2\omega^5kT\theta^2}{c^3\hbar} +
 {\cal O}(\theta^3)\;, \label{thetarj}
\end{eqnarray}
and Wien
\begin{eqnarray}
f_{\theta}(\omega) &=& \frac{8\pi\hbar}{c^3}\cdot{\omega^3}\exp\left(-\frac{\hbar\omega}{kT}\right) \;+\label{thetawien}\\
&-&\frac{8\pi\hbar}{c^3}\cdot\frac{m^2 \omega^5\left(\hbar\omega-kT\right)}{kT\hbar^2} \exp\left(-\frac{\hbar\omega}{kT}\right)\theta^2 + {\cal O}(\theta^3)\;, \nonumber
\end{eqnarray}
distributions respectively.

Let us now find  the  total energy density of radiation by integration of formula
(\ref{planckfunction}) over all frequencies
\begin{equation}
u_\theta = \int_{0}^\infty f_{\theta}(\omega)d\omega
\;; \label{integration}
\end{equation}
the results of numerical calculations are summarized on Figure 3 and 4 respectively\footnote{We use formula (\ref{planckfunction}) with $m=k=c=\hbar=1$ and without factor $8\pi$.}. Consequently one can notice that the value of deformed total radiation energy for fixed temperature $T=1$ and $2$ strongly
decreases with increasing deformation parameter $\theta$. It means that the biggest value of  energy appears for undeformed case
(with $\theta$ equal zero). Such a result (Figure 1-4 and formula (\ref{planckfunction})) formally indicates that the canonical space-time noncommutativity  effectively damps the black-body radiation process.

\section{Final remarks}

In this article we formally investigate the impact of twisted space-time on black-body radiation phenomena. Precisely we derive the $\theta$-deformed Planck distribution function (see formula (\ref{planckfunction}))  as well as we
perform its numerical integration to the $\theta$-deformed total radiation energy. In such a way we indicate that the space-time noncommutativity  very strongly damps the black-body radiation process. Besides we provide for small $\theta$ the twisted counterparts of
  Rayleigh-Jeans and Wien distributions respectively.  Obviously for deformation parameter $\theta$ approaching zero all obtained results become classical.

\section*{Acknowledgments}
The author would like to thank J. Lukierski, W. Sobkow and J. Miskiewicz for valuable discussions.
 This paper has been financially  supported  by Polish
NCN grant No 2014/13/B/ST2/04043.

\eject

\pagestyle{empty}
$~~~~~~~~~~~~~~~~~~$
\\
\\
\\
\\
\\
\\
\\
\\
\\
\begin{figure}[htp]
\includegraphics[width=\textwidth]{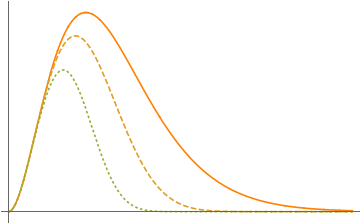}
\caption{The shape of distribution $f_{\theta}(\omega)$ for three different values of
deformation parameter theta: $\theta=0$ (continuous line), $\theta=0.1$ (dashed line) and $\theta=0.2$ (dotted line). In all
three cases we fix temperature $T=1$ and parameter $m=1$.}\label{grysunek1}
\end{figure}
\begin{figure}[htp]
\includegraphics[width=\textwidth]{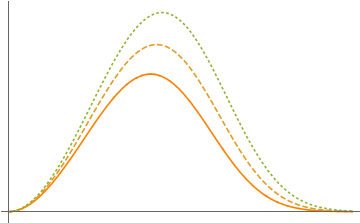}
\caption{The shape of distribution $f_{\theta}(\omega)$ for three different values of
temperature: $T=1$ (continuous line), $T=1.1$ (dashed line) and $T=1.2$ (dotted line). In all
three cases we fix the deformation parameter $\theta=1$ and parameter $m=1$.}\label{grysunek2}
\end{figure}
\begin{figure}[htp]
\includegraphics[width=\textwidth]{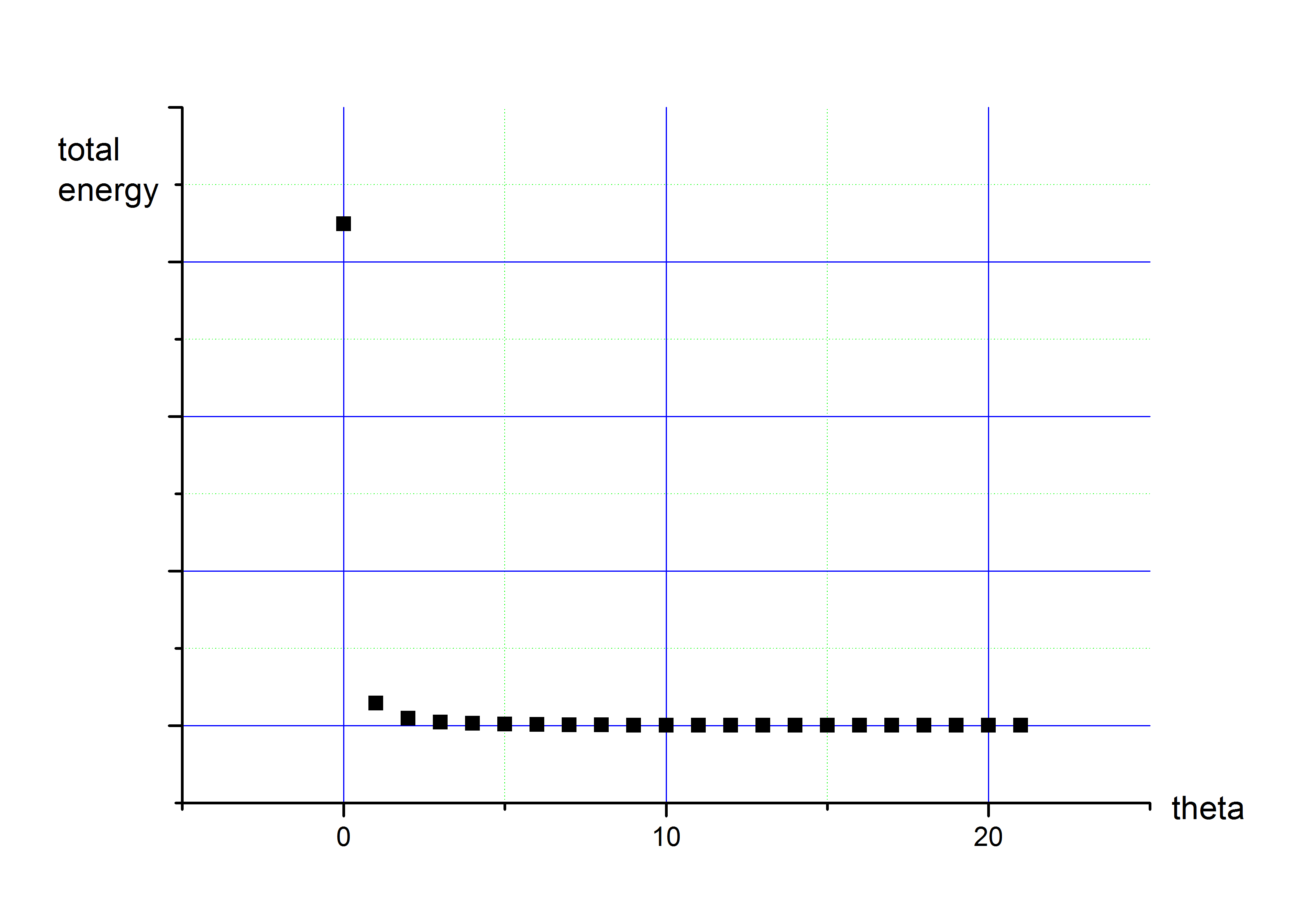}
\caption{The values of total radiation energy $u_{\theta}$ as a function of the
 deformation parameter $\theta=0$, 1, 2, $\dots$, 21 for fixed temperature  $T=1$.}\label{grysunek3}
\end{figure}
\begin{figure}[htp]
\includegraphics[width=\textwidth]{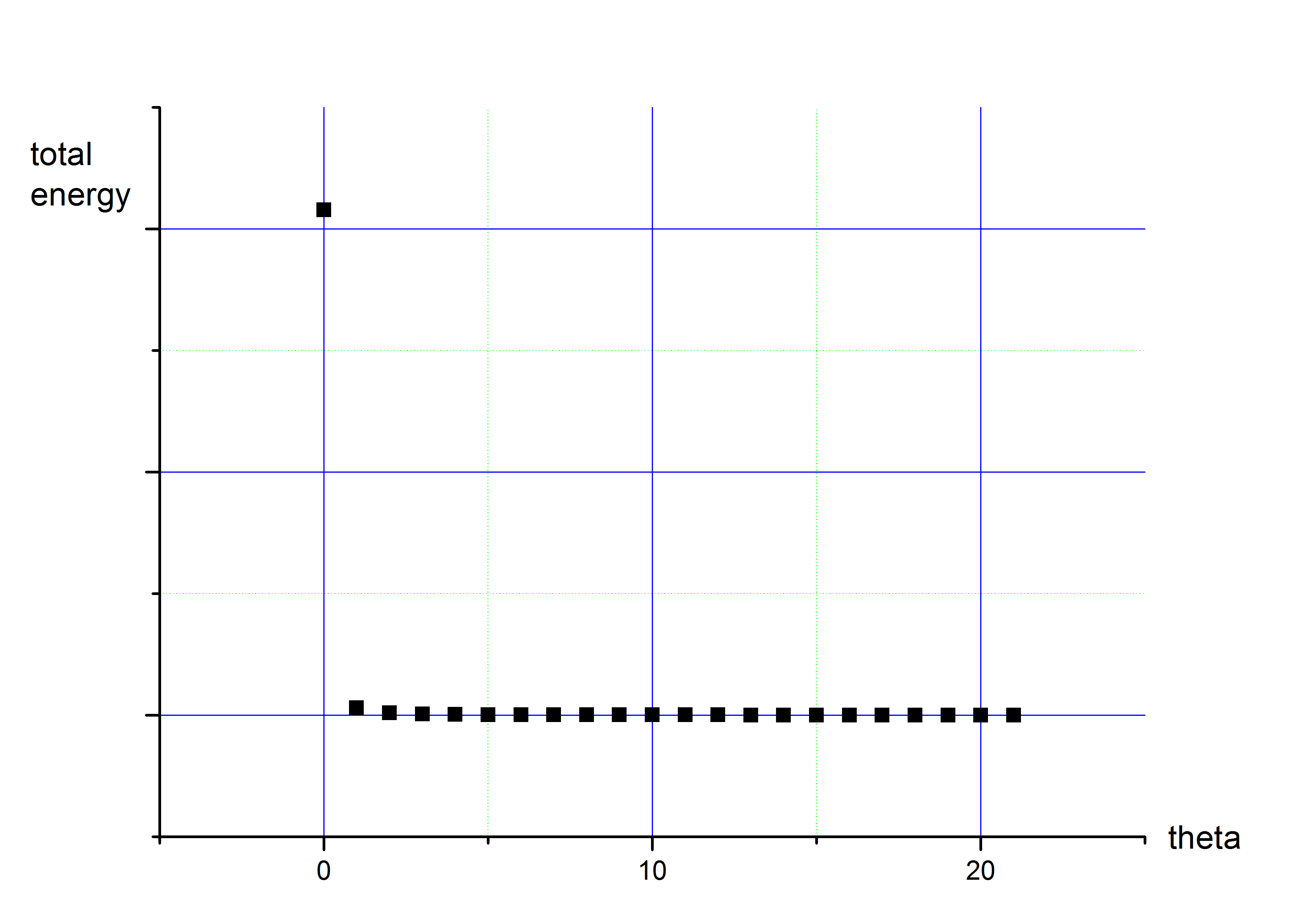}
\caption{The values of total radiation energy $u_{\theta}$ as a function of the
 deformation parameter $\theta=0$, 1, 2, $\dots$, 21 for fixed temperature  $T=2$.}\label{grysunek4}
\end{figure}

\end{document}